\documentstyle[eqsecnum,aps]{revtex}

\begin{document}

\title{Scaling and the Metal-Insulator Transition in Si/SiGe Quantum Wells}

\author{J. Lam}
\address{Dept. of Physics, University of Ottawa, Ottawa, ON, K1N 6N5}

\author{M. D'Iorio , D. Brown and H. Lafontaine}
\address{National Research Council of Canada, IMS, Ottawa, ON, K1A 0R6}

\date{\today}
\maketitle
\begin{abstract}
The existence of a metal-insulator transition at zero magnetic field in two-dimensional electron systems has recently been confirmed in high mobility Si-MOSFETs.  In this work, the temperature dependence of the resistivity of gated Si/SiGe/Si quantum well structures has revealed a similar metal-insulator transition as a function of carrier density at zero magnetic field.  We also report evidence for a Coulomb gap in the temperature dependence of the resistivity of the dilute 2D hole gas confined in a SiGe quantum well. In addition, the resistivity in the insulating phase scales with a single parameter \(T_{0}=\mid n_{c}-n_{s} \mid^{z\nu}\)  where \(z\nu = 1.6\pm0.2\). This dependence is sample independent. These results are consistent with the occurrence of a  metal-insulator transition at zero magnetic field in SiGe square quantum wells driven by strong hole-hole interactions. 
\end{abstract}


\section{Introduction}
\indent For more than two decades, it has been generally accepted that, at zero magnetic field all states are localized in a two-dimensional disordered system as T$\rightarrow$0~\cite{Abr79}.  In contrast, a three-dimensional system can sustain a metal-insulator transition because electrons are localized only when the Fermi energy lies below a mobility edge E$_{c}$.  The conventional wisdom in 2D systems has recently been questioned as a result of a number of experimental results in superconducting thin films ~\cite{Liu91}, and in high-mobility Si-MOSFETs ~\cite{Kra94,Kra95,Kra96,Pop97}. In high mobility Si-MOSFETs, it was recognized that the electron-electron interactions were strong and possibly at the origin of the behavior of the resistivity in the dilute density regime both at zero and non-zero magnetic fields~\cite{DIo92,Pud93}. The most current theories have now focused on the role of strong electron-electron interactions in 2D systems allowing for a metal-insulator transition to occur in 2D as had been suggested by Finkel'stein in the early 1980's~\cite{Fin83}. In cases where the scattering is dominated by the existence of a spin-gap associated with strong spin-orbit coupling, increasing disorder can drive the scaling function from a positive to a negative value thus allowing the metal-insulator transition to take place~\cite{Pud97}.
\medskip
\par
\indent One of the signatures of a phase transition is the scaling of an appropriate physical parameter.  This has recently been demonstrated in high mobility Si-MOSFETs where the scaling conforms with the classical universal exponents~\cite{Kra94,Kra95,Kra96,Pop97}. In these samples, the electron-electron interaction is more than an order of magnitude larger than the Fermi energy. As expected in strongly interacting systems, the temperature dependence of the resistivity for the metallic and insulating phases in Si-MOSFETs is typical of a Coulomb gap~\cite{Mas95}.
\medskip
\par
\indent In this paper, these ideas are tested on another physical system characterized by strong Coulomb interactions. The p-type Si/SiGe material system is interesting because high quality pseudomorphic samples can be grown by ultra-high vacuum chemical vapor deposition (UHV-CVD) with high mobilities and low carrier densities.  In SiGe quantum wells, only valence band holes are significantly confined; the planar biaxial strain resulting from the 4\% lattice constant mismatch between the Si and the Ge and the valence band mixing effects can be used to engineer the electronic properties of the structures. Of particular importance is the substitution of the spin quantum number by parity as a good quantum number because of the mixing of the spin and orbital degrees of freedom in the valence band~\cite{Reg97}. In the symmetric quantum well used for this work, only the ground heavy hole subband is occupied and the resulting charge density is centered in the well. The effective mass of the heavy holes confined in such structures is similar or larger than that of electrons in Si-MOSFETs, \(m_{h}= 0.22-0.44 m_{e}\), and the spin-orbit coupling is large thus setting the stage for cooperative rather than single-particle phenomena.

\section{Samples}
\indent We have studied pseudomorphic Si-Si$_{0.87}$Ge$_{0.13}$-Si heterostructures grown by UHV$-$CVD on an n$-$ Si (100) substrate.  The symmetrically-doped samples were grown on a 500 \AA\ i-Si buffer and consist of a 500 \AA\ Si layer modulation-doped with boron acceptors at 1$\times10^{18}$ cm$^{-3}$ and a 300 \AA\ i-Si setback on each side of a 65 \AA\  Si$_{0.87}$Ge$_{0.13}$ quantum well.  The cap layer consists of 395 \AA\ of B-doped Si.  The samples were patterned as Hall bar structures 10 mm long and 2 mm wide, and ohmic contacts were made using Al evaporation and annealing below the eutectic point.  A 900 \AA\ Ti-Au Schottky gate was deposited on top of the cap layer.  Without bias applied to the gate, the mobility and carrier density at 1.7 K are 1100 cm$^{2}/Vs$ and 2.6$\times10^{11}$ cm$^{-2}$ respectively.  In this study the carrier density was varied from 1.60$-$4.3$\times10^{11}$ cm$^{-2}$ with bias voltages in the range +200 mV to $-50$ mV. The bias versus carrier density variation is linear in the high density regime and the linear dependence is extrapolated to lower densities.  Measurements of the metallic phase above 4.3$\times10^{11}$ cm$^{-2}$ have not been carried out because of leakage of the Schottky gate. 
\medskip
\par
\indent Previous optical and electronic transport measurements~\cite{DIo96a,DIo96b,Col97} of the effective mass in identical samples grown by MBE suggest that the effective mass is of the order of $0.4m_{e}$ and that the holes are confined in the ground heavy hole subband. The quantum mobilities determined by the low field Shubnikov-de Haas oscillations are similar to the transport mobilities as would be expected for large angle scattering from impurities close to the interface~\cite{Lam97,Eme93,Bas94}. The samples were measured in the temperature range 25 mK-4.2 K, using four-terminal dc techniques and a high input impedance differential electrometer.

\section{Results and discussion}
\indent As shown in Figure 1, the resistivity in the high density regime undergoes a monotonic decrease with decreasing temperature corresponding to a reduction of phonon scattering. The precipitous drop of the resistivity below 2K observed in Si-MOSFETs and in wider SiGe quantum wells~\cite{DIo96a,DIo96b,Col97} is not observed in these samples as gate leakage in the bias range corresponding to higher carrier densities made such measurements impossible. The temperature dependence of another 65 \AA\ SiGe quantum well is shown as an inset to Figure 1 and suggests that the metallic phase in SiGe quantum wells may well be very similar to that observed in high mobility Si-MOSFETs. Near a critical density \(n_{c} = 2.2\times10^{11}\) cm$^{-2}$, the resistivity in the gated sample starts to rise with decreasing temperature signaling the onset of the metal-insulator transition. Figure 2 shows the temperature dependence of the resistivity above 400 mK as a function of carrier density.  A best fit to these curves indicates that the temperature dependence can be described as $\rho_{c}exp(T_{0}/T)^{1/2}$ for carrier densities lower than $n_{s} = 1.9\times10^{11}$ cm$^{-2}$ where $\rho_{c}$ is approximately $h/e^{2}$, the quantum of resistivity. This behavior is observed over more than an order of magnitude in resistivity and is characteristic of conduction by hopping in the presence of a Coulomb gap. Using the Efros and Shklovskii model of Coulomb interaction between localized carriers~\cite{Efr75}, the characteristic temperature $T_{0}$ can be expressed as:
 \[k_{B}T_{0}=\frac{Ce^{2}}{\epsilon \xi}\]
where $\xi$ is the localization length, $\epsilon$ is the dielectric constant, $e$ is the charge of the carriers and $C$ is a constant taken to be $\approx 6.2$. The Coulomb gap behavior is observed down to 400 mK while the characteristic temperature $T_{0}$ spans the range 0.3-1.5 K for a localization length variation of 28-6 mm. The effective screening radius in the gated SiGe samples is d=80 nm, defined by the distance between the 2D holes and the gate. It is much smaller than the localization length. While the Coulomb gap behavior would be expected only when the hopping length $(1/4\xi(T_{0}/T)^{1/2})$ is shorter than the screening radius, it is clear that the hopping length is more than an order of magnitude larger than 2d. Such discrepancies were also observed in Si-MOSFETs and could be reconciled only if the value of the coefficient C is much smaller.
\medskip
\par
\indent More striking is the fact that the data may be scaled such that it collapses unto one curve through use of the single scaling parameter T$_{0}$ in the temperature dependence of the resistivity: $\rho(T,n_{s})= \rho[T/T_{0}(n_{s})]$.  T$_{0}^{(1)}$ was selected from the fit to the lowest density curve, and then applied to all subsequent curves such that $T_{0}^{(i)}=\gamma^{(i)}T_{0}^{(1)}$.  Figure 3 shows the collapsed data, scaled by the factors $\gamma(i)$ to coincide with the most insulating curve with each curve representing a different carrier density.  The relationship of the scaling parameter to the carrier density is given by $T_{0}((\delta_{n})=A\mid(\delta_{n})\mid^{\beta}$ , where $\delta_{n}= (n_{c}-n_{s})$, $n_{c}$ is the critical carrier density at which the transition occurs, and $\beta = z\nu$ where $\nu$ is the correlation length exponent and $z$ is the dynamical exponent. The power law relation for T$_{0}$ is exactly satisfied for $n_{c}=2.2\times10^{11}$ cm$^{-2}$ and all the data collapses on a single curve for the insulating side of the transition. The best fit was obtained for $z\nu =1.6 \pm 0.2$ with a carrier density variation up to $\delta_{n}=0.31 \times 10^{11} cm^{-2}$. This agrees with both previous and more recent results for high mobility Si MOSFETs. The inset in Fig. 3 shows the collapsed data scaled along the temperature axis as $(T_{0}/T)^{1/2}$. The curve extrapolates to the quantum of resistivity $h/e^{2}$ and deviates from the Coulomb gap behavior when $T\approx T_{0}$.
\medskip
\par
\indent In summary we have shown that the zero magnetic field resistivity of 2D holes in high quality SiGe quantum wells scales with temperature. Below a critical carrier density $n_{c}$, a single scaling parameter can be used to collapse the resistivity on a single curve in the insulating phase. The metallic phase was not studied in this work but is known from other measurements to behave in a similar fashion. The critical exponent is $1.6\pm 0.2$ and the temperature dependence of the resistivity is consistent with the opening of the Coulomb gap as is expected for a system with interacting carriers. The results are consistent with those obtained in high mobility Si-MOSFETs and  suggest that the metal-insulator transition at zero magnetic field in p-type SiGe quantum wells is driven by the Coulomb interaction. 

\medskip
\par
Acknowledgments
\medskip
\par
\indent We would like to thank Duncan Stewart (Stanford Un.)  and Simon Deblois (Un. Laval) for their skilled contributions in sample design, fabrication, processing, characterization and modeling. Graduate student support  was provided by an operating grant from the Natural Sciences and Engineering Research Council (NSERC) and by the National Research Council.

\medskip
\par
Figure Captions
\medskip
\par
Fig. 1: The temperature dependence of the resistivity in the UHV-CVD grown 65 \AA\ square quantum well (CVD32K) in the insulating phase and close to the metal-insulator transition. The inset shows the temperature dependence of another 65 \AA\ square quantum well (CVD32J) where the behavior in the metallic phase is better resolved.
\medskip
\par
Fig. 2: The temperature dependence of the resistivity in the insulating phase for sample CVD32K in the temperature range 0.5-2 K.
\medskip
\par
Fig. 3: Scaling of the resistivity as a function of $(T/T_{0})$. The inset shows the resistivity as a function of $(T_{0}/T)^{1/2}$ for CVD32K.

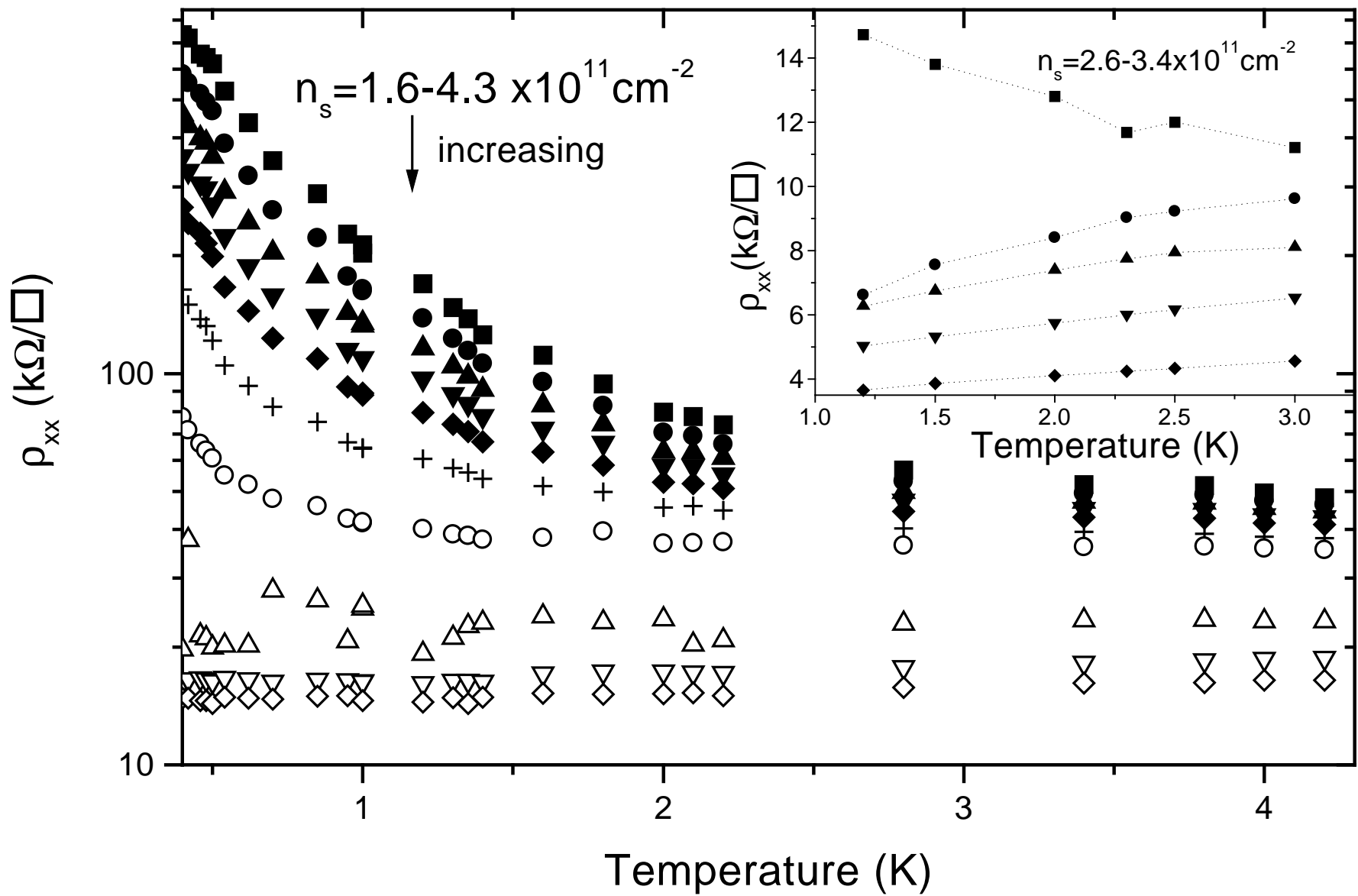
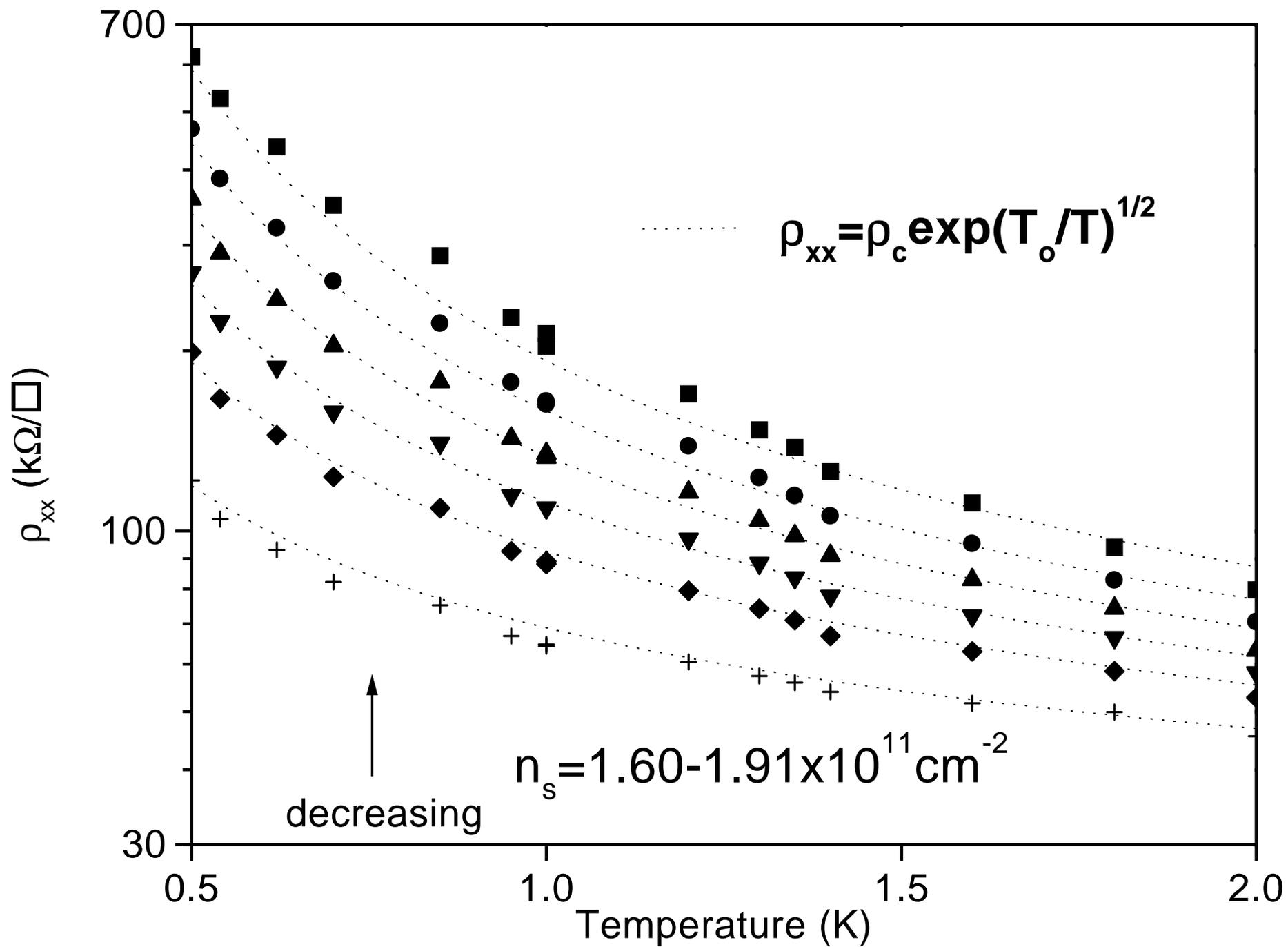
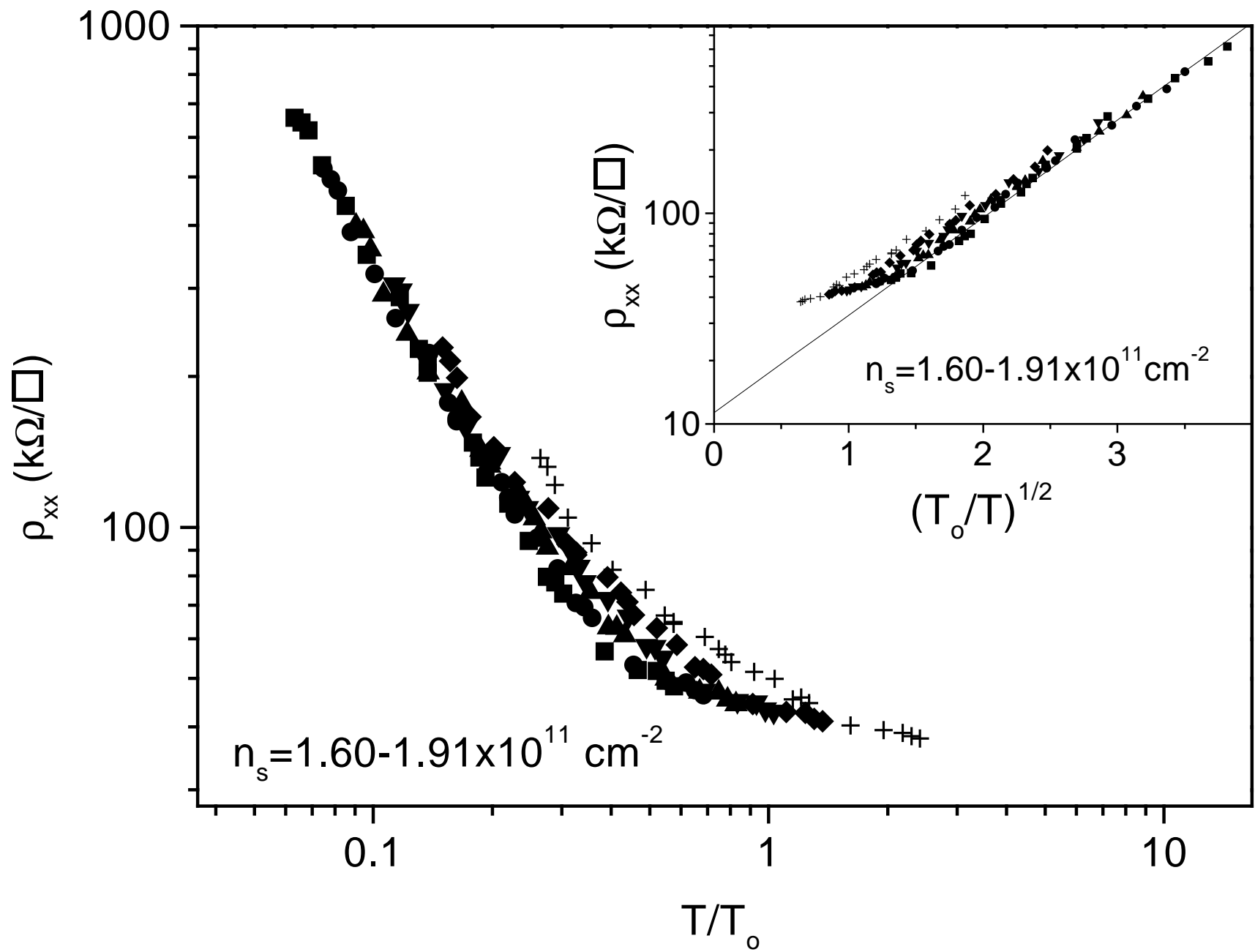
\end{document}